\documentclass[journal]{IEEEtran}
\usepackage{booktabs} 
\usepackage{graphicx}
\usepackage{epstopdf}
\DeclareGraphicsExtensions{.eps}
\usepackage{amsmath}
\usepackage{bm}
\usepackage{float}
\usepackage{url}
\usepackage{balance}
\usepackage{subcaption}

\usepackage{setspace}

\ifCLASSINFOpdf
\else
\fi
\begin{document}
%
\title{CloudAR: A Cloud-based Framework for Mobile Augmented Reality}
%
%
%

\author{Wenxiao ZHANG,
        Sikun LIN,
        Farshid Hassani Bijarbooneh,
        Hao Fei CHENG,
        and~Pan HUI,~\IEEEmembership{Fellow,~IEEE}
\thanks{W. ZHANG is with the Department of Computer Science and Engineering,
the Hong Kong University of Science and Technology, Hong Kong. E-mail: wzhangal@cse.ust.hk}
\thanks{S. LIN is with the Department of Computer Science,
University of California, Santa Barbara. E-mail: sikun@ucsb.edu}
\thanks{F. H. Bijarbooneh is with the Department of Computer Science and Engineering,
the Hong Kong University of Science and Technology, Hong Kong. E-mail: farshidhss@ust.hk}
\thanks{H. F. CHENG is with the Department of Computer Science and Engineering,
University of Minnesota, U.S.. E-mail: cheng635@umn.edu}
\thanks{P. HUI is with the Department of Computer Science and Engineering,
the Hong Kong University of Science and Technology, Hong Kong, and the Department of Computer Science, University of Helsinki, Finland. E-mail: panhui@cse.ust.hk}
}

\maketitle

\begin{abstract}
Computation capabilities of recent mobile devices enable natural feature processing for Augmented Reality (AR). However, mobile AR applications are still faced with scalability and performance challenges. In this paper, we propose CloudAR, a mobile AR framework utilizing the advantages of cloud and edge computing through recognition task offloading. We explore the design space of cloud-based AR exhaustively and optimize the offloading pipeline to minimize the time and energy consumption.
We design an innovative tracking system for mobile devices which provides lightweight tracking in 6 degree of freedom (6DoF) and hides the offloading latency from users' perception.
We also design a multi-object image retrieval pipeline that executes fast and accurate image recognition tasks on servers.
In our evaluations, the mobile AR application built with the CloudAR framework runs at 30 frames per second (FPS) on average with precise tracking of only 1$\sim$2 pixel errors and image recognition of at least 97\% accuracy. Our results also show that CloudAR outperforms one of the leading commercial AR framework in several performance metrics.
\end{abstract}


\section{Introduction}
Augmented Reality (AR) is a natural way of interaction between the real world and digital virtual world. For a typical AR application, it recognizes the surrounding objects or surfaces and overlays information on top of the camera view with a 3D renderer. Currently, mobile Augmented Reality (MAR) is the most practical AR platform as mobile devices are widely used in our daily life, and many mobile AR SDKs (e.g., Apple ARKit \cite{arkit}, Google ARCore \cite{arcore}, Vuforia \cite{vuforia}) are released to enable fast development of AR Apps.
However, mobile devices are still suffering from the inherent problems of mobile platforms (e.g., limited computation power and battery life), which restrict their performance in practical AR applications.
Some works \cite{hagbi2009shape,wagner2010real, wagner2009multiple} have showed the recognition, tracking, and rendering capabilities of mobile devices, nevertheless, like most of the AR applications in the market, they are working in simple scenarios, displaying some fixed contents based on detected surfaces, and limiting their usage in gaming or simple demonstration. 

The key enabler of practical AR applications is context awareness, with which AR applications recognize the objects and incidents within the vicinity of the users to truly assist their daily life.
Large-scale image recognition is a crucial component of context-aware AR systems, leveraging the vision input of mobile devices and having extensive applications in retail, education, tourism, advertisement, etc.. For example, an AR assistant application could recognize road signs, posters, or book covers around users in their daily life, and overlays useful information on top of those physical images.

Despite the promising benefits, large-scale image recognition encounters major challenges on the mobile platform. First, large-scale image recognition requires the storage of large image dataset, and the corresponding annotation contents are also huge in scale and size. Second, the image recognition task itself is computation intensive and consumes a lot of time and energy \cite{yang2015mobile}. According to a study \cite{shoaib2015exploiting}, it takes on average longer than 2 seconds to finish object recognition on a mobile CPU (Snapdragon 800). 

\begin{figure}[t!]
\center 
\begin{subfigure}[t]{0.40\textwidth}
    \includegraphics[width=\textwidth]{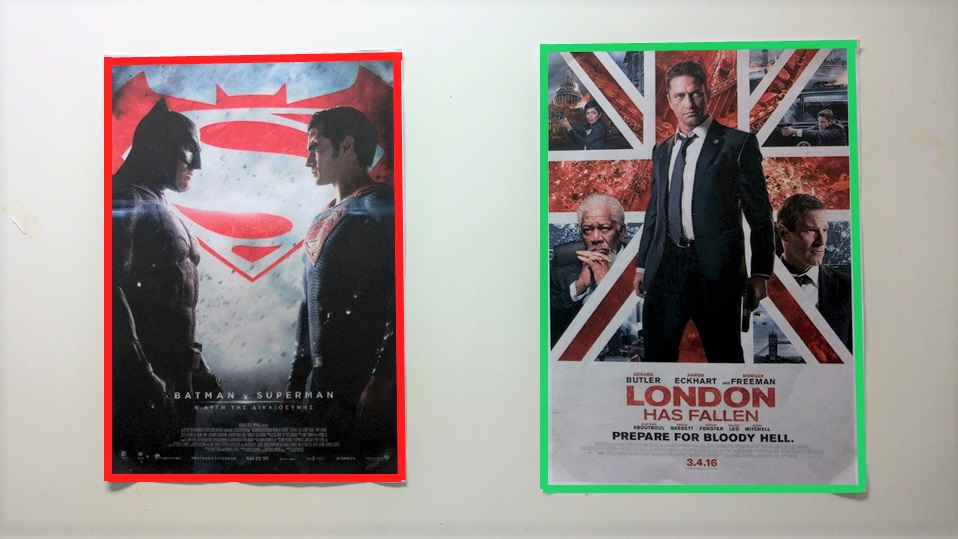}
    \caption{Server recognizes images in the request frame.}
    \label{fig:request}
    \end{subfigure}
\begin{subfigure}[t]{0.40\textwidth}
    \includegraphics[width=\textwidth]{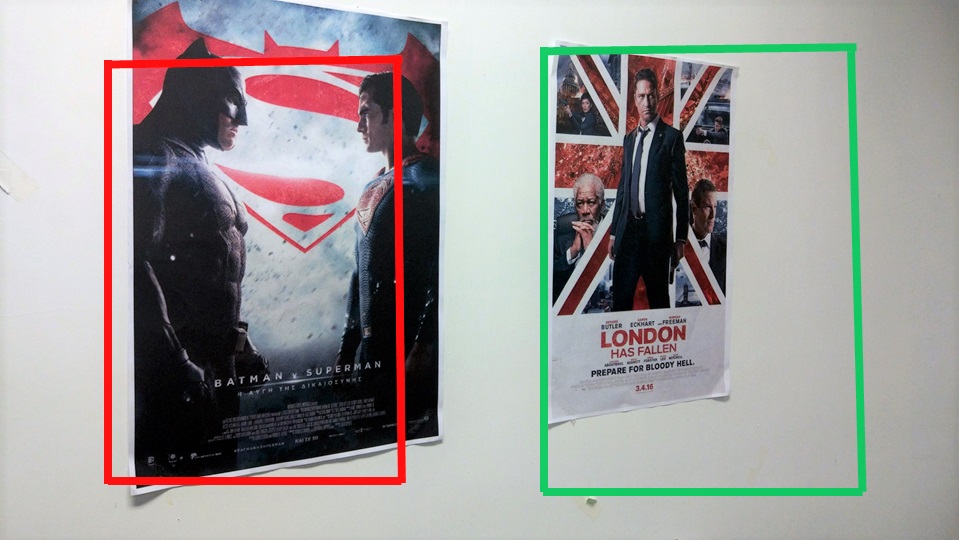}
    \caption{Results mismatch with the current view.}
    \label{fig:result}
    \end{subfigure}
\caption{Mismatching problem in cloud-based AR caused by cloud recognition latency.}
\label{fig:mismatch}
\end{figure}

By utilizing cloud image recognition, cloud-based AR systems\footnote{We use \emph{cloud-based system} to refer to both edge servers and cloud servers.} \cite{pilet2010virtually,gammeter2010server,jung2012efficient} are able to fill the gap between context-aware AR experience and insufficient mobile capabilities.
Similar to web applications, context-aware AR applications should communicate with the cloud to offload the image recognition tasks and retrieve useful information. 
However, cloud-based systems suffers from the offloading latency, which consists of both the network transferring delay and server processing time. Results of the recognition requests are always returned after a moment, during which the users are likely to move their hands. Unlike typical web applications, AR applications have to align the recognition results precisely with the physical world. The cloud recognition latency makes the results mismatch with the current view, as shown in Figure \ref{fig:mismatch}, and damages the quality of experience (QoE) significantly. 

With the inevitable offloading latency, mobile AR applications should handle the outdated result properly. Existing cloud-based AR systems fail in solving this mismatching problem. They either ask the users to hold their hands steady to coarsely match the result with the current view, or neglect the performance constraints and real-time interaction requirement of mobile devices. 

To address the scalability and latency issues in MAR applications, we propose CloudAR, a cloud-based MAR framework with both an innovative mobile client design and a powerful image recognition pipeline running on cloud servers. One major advantage of CloudAR is that it solves the mismatching problem with an innovative tracking system running on the mobile client, where the historical tracking status of the request frame is recorded and finally utilized by the tracker to compute a corrected position of the physical image after receiving the result. With our fine-grained AR offloading pipeline, AR applications developed with CloudAR support large-scale image recognition with the guarantee of high QoE.
We achieve this with the following contribution:
\begin{itemize}
\item A lightweight mobile tracking system that provides precise 6 degree-of-freedom (6DoF) multi-object tracking, hides mismatching of the recognition result from user's perception, and achieves real-time performance at 30 frames-per-second (FPS).
\item A multi-object cloud image recognition pipeline that provides low-latency, high-accuracy multi-image recognition, as well as pose estimation services.
\item A fine-grained cloud offloading pipeline that minimizes the overall recognition latency and mobile energy consumption, and a systematic evaluation with comparison to a popular commercial MAR SDK.
\end{itemize}

The remainder of this paper is organized as following:  Section~\ref{sec:relatedwork} introduces the existing work on MAR systems and latency compensation techniques. Section~\ref{sec:designchoice} discusses the design choices of CloudAR. Section~\ref{sec:systemDesign} gives an overview of the system architecture. Section~\ref{sec:clientDesign} and~\ref{sec:serverDesign} describe the mobile client design and server design, respectively. The implementation details and the experimental results are shown in Section~\ref{sec:evaluation}. Finally, the conclusion is given in Section~\ref{sec:conclusion}.

\section{Related Work}
\label{sec:relatedwork}

\subsection{On-device mobile Augmented Reality}
To precisely overlay annotation contents on top of the physical environment, a typical AR application calculates the physical camera pose and aligns the renderer's virtual camera accordingly for each frame, where the pose is in 6 degree of freedom (3 degree of translation and 3 degree of rotation). The derived camera pose can be either relative to an image, or absolute in the environment. 

Typical mobile AR SDKs can be divided into two categories: a) traditional SDKs (e.g., Wikitude \cite{wikitude}, Vuforia \cite{vuforia}, Blippar \cite{blippar} with their basic functions) that heavily utilize computer vision technologies for marker-based AR, which recognize a marker first and track the relative camera pose to the marker for each frame; b) emerging AR SDKs (e.g., Google ARCore \cite{arcore}, Apple ARKit \cite{arkit}) that bring marker-less AR onto the mobile platform, which calculate an absolute pose of the camera in the environment with visual-inertial odometry (VIO) \cite{li2013high,leutenegger2015keyframe}.

For marker-based AR, Nate et al.~\cite{hagbi2009shape} proposes a mobile system to recognize and estimate 6DoF pose of planar shapes, and applies recursive tracking to achieve interactive frame rate.
Wagner et al. modifies SIFT and Ferns to make them suitable for mobile platforms~\cite{wagner2010real}, and builds a system to detect and track multiple objects on a mobile phone~\cite{wagner2009multiple}. Global object pose estimation and device localization are also implemented on the mobile platform \cite{ventura2014global,arth2015global}.

For marker-less AR, both ARCore and ARKit leverage feature points from the monocular camera and motion data from the inertial measurement unit (IMU) to track the pose of the mobile device in 3D. Their tracking capability is more flexible and scalable compared to marker-based AR. 

\subsection{Cloud-based Mobile Augmented Reality}
Google Goggles \cite{goggles} lets a user search by taking a picture. If Goggles finds it in its database, useful information will be provided. 
The systems in \cite{pilet2010virtually,gammeter2010server} try to address the scalability problem and integrates the tracking system with image retrieval techniques on a PC, but they cannot handle the mismatching problem.
A similar system is described in \cite{jung2012efficient}, where extra information like region of interest (ROI) of the object is required from the user to initialize the recognition and tracking. 
Overlay \cite{jain2015overlay} also utilizes server for image retrieval, and they requires the user to hold the camera still to let annotation display in a coarsely correct position.
VisualPrint \cite{jain2016low} is another cloud-based mobile augmented reality system which uploads extracted features instead of raw images or video stream to save network usage. 

Compared to existing cloud-based AR systems, the CloudAR framework handles the offloading latency properly and practically, and minimizes the overall latency and energy consumption.

\subsection{Image Retrieval Techniques}
To detect and extract features from an image, SIFT \cite{lowe1999object}, and SURF \cite{bay2006surf} provide good results with robust scale and rotation invariant, and CNN features \cite{krizhevsky2012imagenet} provide state-of-the-art representation accuracy. 
However, these approaches are slow. Combined with binary descriptors, corner feature points detectors such as Harris corner detector \cite{harris1988combined}, FAST \cite{rosten2010faster}, and AGAST \cite{mair2010adaptive} are faster choices.

To further retrieve or classify images, one image can be encoded into a single vector. The authors in \cite{csurka2004visual} introduce the bag-of-visual-words (BOV) model.
The authors in \cite{jaakkola1999exploiting,perronnin2010improving,jegou2012aggregating} present Fisher kernel encoding as a statistically more accurate representation than k-means clustering used in BOV.
In recent years, Fisher Vector (FV) is widely used in image retrieval with good performance as shown in \cite{perronnin2010large,douze2011combining}. 
The authors in \cite{indyk1998approximate,gionis1999similarity} propose LSH for fast nearest neighbor search, and many improvements are proposed in \cite{datar2004locality,andoni2006near,kulis2009kernelized}. 

\subsection{Latency Hiding Techniques}
Volker Strumpen et al. presentes a latency hiding protocol for asynchronous message passing in UNIX environments \cite{Strumpen1995}. With this protocol distributed parallel computing can be utilized in applications.

Outatime \cite{lee2015outatime} is a mobile cloud gaming system which delivers real-time gaming interactivity. The basic approach combines input prediction with speculative execution to render multiple possible frame outputs which could occur in a round trip time of future. 
Kahawai \cite{cuervo2015kahawai} is another mobile gaming systems which aims at providing high-quality gaming quality by offloading a portion of the GPU computation to server-side infrastructure. In contrast with previous thin-client approaches which require a server-side GPU to render the entire content, Kahawai uses collaborative rendering to combine the output of a mobile GPU and a server-side GPU into the displayed output. 

TYH Chen et al. \cite{chen2015glimpse} tries to solve the offloading latency problem on mobile by caching frames during the offloading procedure and running visual tracker on the frames upon receiving the result. Up-to-date position of the objects can be found with their approach, but this method may cause instant computationally intensive demands. Meanwhile, their object tracking system is 2D based which cannot give the 6DoF poses of objects.

Different from those existing methods, CloudAR targets specifically at latency compensation under the scenario of AR, which is more challenging as the latency problem is more sensitive with the alignment of virtual and physical world and the physical environment is less predictable. CloudAR provides seamless real-time user experiences while keeping the mobile client as lightweight as possible.

\section {Background and Design Choices}
\label{sec:designchoice}

\begin{figure*}[t]
    \centering
    \includegraphics[width=1\textwidth]{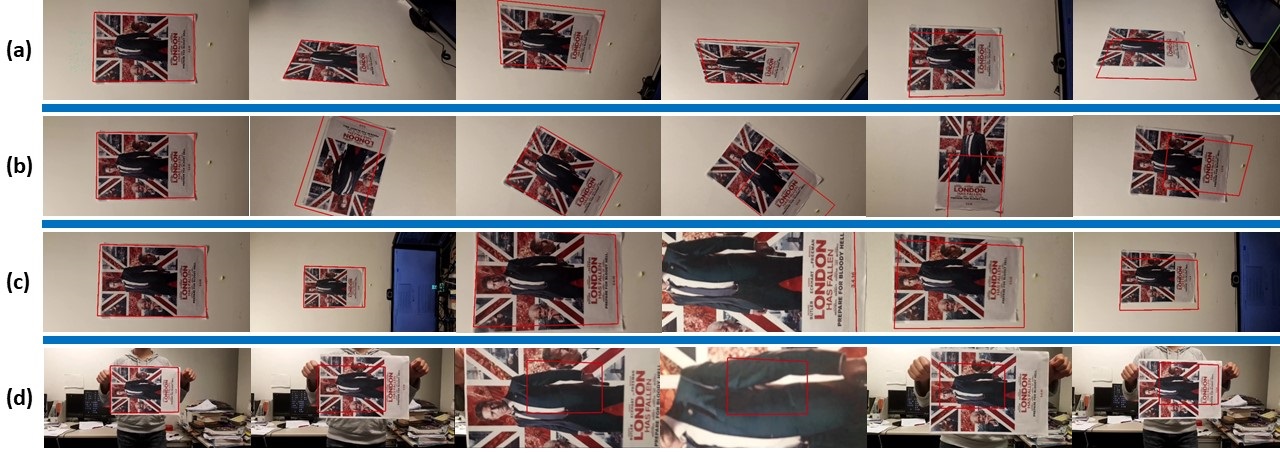}
    \caption{Results of ARCore tracking in four scenarios: (a) view changing; (b) camera rotation; (c) scaling caused by forward and backward movement; (d)object moving. The red edges are the tracking results on the mobile client.}
    \label{fig:arcore}
\end{figure*}

In this section, we explore the design space of the cloud-based AR systems, and discuss the advantages and disadvantages of possible solutions to motivate the design of CloudAR.

\subsection{Is VIO a good solution for context-aware AR?}

ARKit and ARCore hit the market, and the technology underneath, VIO locates mobile devices in the environment independent of the detection of specific markers, providing a more flexible and scalable tracking capability when compared with traditional marker-based AR. However, VIO cannot recognize the surrounding objects and get context information. Most of existing AR Apps powered by ARCore or ARKit  work on the detected horizontal surfaces (e.g., table, floor) for gaming or demonstration.

To learn the tracking performance of ARCore, especially in the scenario of tracking recognized image, we build an AR App based on ARCore, which recognizes one image stored in the App and tracks it afterwards. The App is deployed on a Samsung Galaxy S8 smartphone, which is a flagship model and well-calibrated for ARCore. The tracking results of four representative sequences are shown in Figure \ref{fig:arcore}, from which we find several problems of VIO-based AR systems:

\textbf{VIO tracking of markers are inaccurate and unstable.} As shown in Figure \ref{fig:arcore}, ARCore tracking results are not accurate in the cases of camera tilt, rotation and scaling, with obvious mismatching between the physical image and the tracking result. Unstable tracking results arise frequently as well, where the annotated image edges hop during tracking as shown in the second half of sequence (b), and this is not a rare case as hopping happens in four out of ten repeated attempts. As VIO tracking combines both the visual and inertial outputs, the noisy and biased inertial measurements are the main reason for this unsatisfying performance.

\textbf{VIO needs good initialization.} Upon loading of the App, a initialization procedure is needed before the tracker would work. The initialization usually takes few seconds, and the user has to hold and move the phone for a few seconds. On the contrary, visual-based tracker used in marker-based AR would start working instantly upon loading without an initialization.

\textbf{VIO cannot track moving objects.} Unlike the tracking of object's relative pose in visual tracking systems, typical VIO systems are supposed to work in a rigid environment, where the localization procedure locates the devices' pose and ignores moving objects, as shown in sequence (d) of Figure \ref{fig:arcore}. Although some existing work have addressed the moving object tracking problem \cite{wang2003online,wang2007simultaneous} in simultaneous localization and mapping (SLAM)  systems, these methods make major modifications of the SLAM pipeline, which add up to the overall complexity and is not currently supported by ARCore or ARKit.

\textbf{VIO needs physical size of recognized image.} Without considering the dynamic object tracking, the static object tracking of VIO is different from that of visual tracker as well. Visual tracker works with the relative pose and size of the physical object and annotation contents, so that the cloud recognition service would return the pose of the image object without caring about the real size of the image. However, as the odometry unit of VIO systems is in actual physical size, the tracking of static objects needs the physical size of the image as well, which is not possible to infer on the cloud server and not trivial to get on the mobile client. In this App developed with ARCore, we have to manually input the physical size of the image, which limits its pervasive usage. 

With the shortages shown above, a VIO based solution is not adequate in context-aware AR with image recognition functionalities. Marker-based AR performs better in terms of accuracy, robustness, ease of use, and availability for image tracking at the current stage.

\subsection{Marker-based AR pipeline}

\begin{figure}[t]
    \centering
    \includegraphics[width=0.5\textwidth]{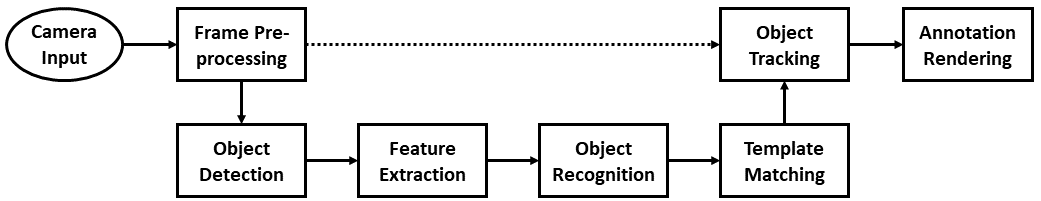}
    \caption{Typical mobile AR pipeline.}
    \label{fig:arpipeline}
\end{figure}

A typical pipeline of marker-based mobile AR systems has 7 building blocks, as shown in Figure \ref{fig:arpipeline}. It starts with Frame Preprocessing that shrinks the size of a camera frame, e.g., by downscaling it from a higher resolution to a lower one. The next step is Object Detection that checks the existence of targets in the camera view of a mobile device and identifies the regions of interest (ROI) for the targets. It will then apply Feature Extraction to extract feature points from each ROI and Object Recognition to determine the original image stored in a database of to-be-recognized objects. Template Matching verifies the object-recognition result by comparing the target object with the recognized original image. It also calculates the pose of a target (i.e., position and orientation). Object Tracking takes the above target pose as its initial input and tracks target object between frames in order to avoid object recognition for every frame. Finally, Annotation Rendering augments the recognized object by rendering its associated content.

\subsection{Latency Compensation Choices}

\begin{table}
  \begin{tabular}{lll}
    \toprule
    Template Matching & Mobile Client & Cloud Server  \\
    \midrule
    Time Consumption(ms) & $438.2\pm96.5$ & $57.62\pm6.14$\\
    Energy Consumption(uAh) & $276.0\pm60.9$ & N/A \\
    \bottomrule
  \end{tabular}
\caption{Time and energy consumption for template matching with SIFT features on a Xiaomi Mi5 phone and a cloud server, respectively. The value after $\pm$ is the standard deviation (for 20 experimental runs).}
\label{tab:templatematching}
\end{table}

As the cloud recognition latency is inevitable, cloud-based AR systems need to handle this latency properly so that the annotation contents can be aligned with the physical objects precisely. For marker-based AR, there are two ways to hide the latency.

One possible solution is that the server sends the original image back and the client executes template matching to calculate the current pose of the image. However, template matching is not a trivial task on mobile, which consumes much time and energy when compared with cloud execution, as shown in Table \ref{tab:templatematching}. Executing template matching on the server reduces the overall pipeline latency significantly. On the other hand, even if template matching is executed on the client, the time consumption of template matching would also make the mismatching happen, as several frames have passed during the template matching procedure.

Another solution is tracking the image movement during the cloud recognition procedure, as tracking is much more light-weight compared to template matching. With the tracked image pose transformation during the cloud recognition procedure, the recognition result can be corrected accordingly to match the current view. A previous work Glimpse \cite{chen2015glimpse} proposes to solve the mismatching problem with tracking as well, but Glimpse caches selected frames during cloud recognition and starts tracking after receiving the result, with the recognition result utilized as the tracking initial input. However, this design would erupt processing burden on the mobile device, and the tracking of a sequence of frames takes long time and causes mismatching as well. 

As a result, our system disperse the tracking processing during the cloud recognition procedure. Our design achieves overall low latency while keeping the mobile client lightweight and fluent.

\section{System Design Overview}
\label{sec:systemDesign}

\begin{figure}[t]
    \centering
    \includegraphics[width=0.5\textwidth]{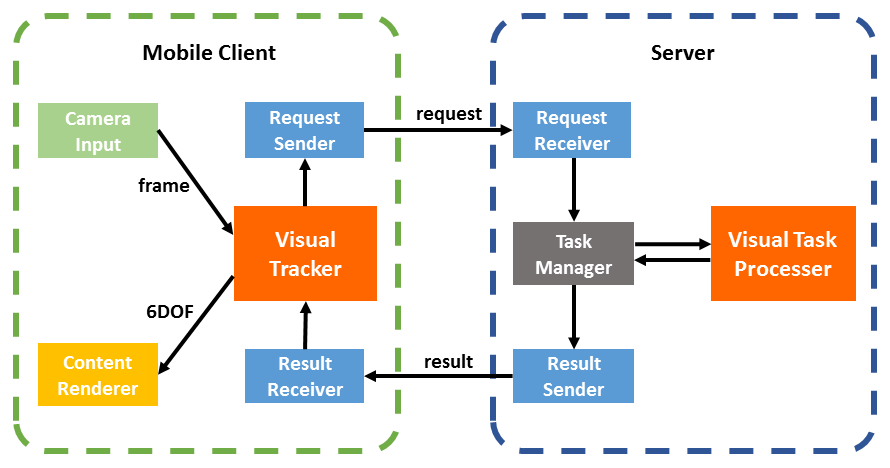}
    \caption{CloudAR overview showing the data flow and the main components of the mobile client and the server.}
    \label{fig:overview}
\end{figure}
Figure \ref{fig:overview} shows an overview of the system architecture, that includes the mobile client and the cloud server. 
Following is a brief description of the overall procedure: 

On the mobile client, visual tracker gets the camera video feed and starts extracting and tracking feature points within view once the application is launched. Unlike \cite{jung2012efficient}, our system would not require the user to input the region of interest (ROI), and camera stream is the only input of the whole framework. Therefore, the visual tracker simply tracks feature points of the entire frame at this time. An object recognition request is sent to server to recognize target objects inside the camera frame. 

The server creates worker threads to process the received request. These worker threads handle the receiving and sending messages on the network, as well as performing the visual tasks. 

The result is then sent back to the mobile client, including poses and identity information of the recognized objects. With feature points tracking, visual tracker is able to calculate the location transition of any specific area inside the frames. Therefore the current poses of the objects can be derived from the result, which are poses of objects within the request frame. These poses are further utilized by the content renderer to render virtual contents. 

This process repeats periodically under visual tracker's scheduling, thus new objects are recognized and augmented continuously.

\section{Mobile Client Design}
\label{sec:clientDesign}
The mobile client mainly works on tracking and contents rendering, with the heavy image recognition tasks offloaded to the cloud. Since the main working flow is already described in Section~\ref{sec:systemDesign}, here we focus on introducing these functional modules on mobile client and their working principles. 

\subsection{Visual Tracker}

\begin{figure}[t]
    \centering
    \includegraphics[width=0.5\textwidth]{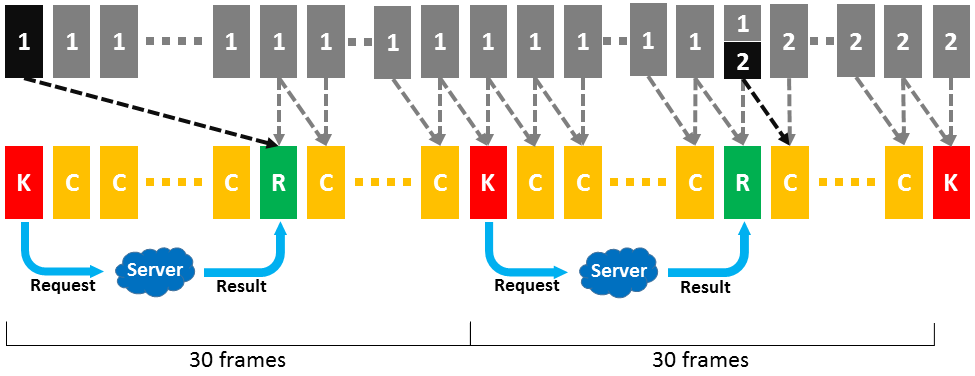}
    \caption{Visual tracker pipeline. The sequence above in grayscale stands for the feature point extraction and tracking pipeline, where each number indicates the feature points set number; the sequence below in color stands for the objects pose initialization and update pipeline, where K stands for a key frame, C stands for a continuous frame, and R stands for a result frame.}
    \label{fig:frames}
\end{figure}

\begin{figure*}[t]
    \centering
    \includegraphics[width=1\textwidth]{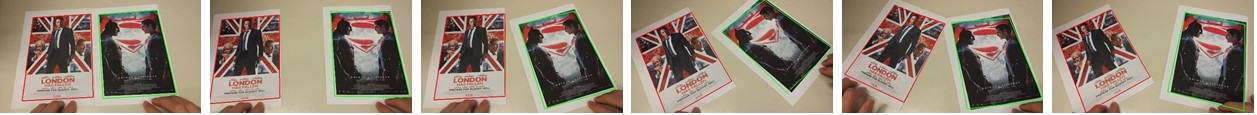}
    \caption{Multi-object tracking of CloudAR. The red and green boundaries are the output of the visual tracker.}
    \label{fig:multitrack}
\end{figure*}

Visual tracker carries out the core logic of mobile client, scheduling the execution of other modules. We design our own feature point based tracking method for mainly two reasons. First, in CloudAR tracking starts without a ROI or the object position, and the renderer needs precise 6DoF results rather than simple rectangle bounding boxes to position virtual contents in a 3D space. Feature point based tracking gives us the flexibility to do these calculations. Second, visual tracker must achieve a high FPS on the mobile client, and feature point based tracking (e.g., optical flow \cite{lucas1981iterative}) is lightweight enough to fulfill this requirement.

We decouple the feature tracking and object pose estimation tasks into two separate pipelines, as shown in Figure \ref{fig:frames}. The first pipeline works on feature point extraction and tracking for the whole frame, while the second pipeline works on pose initialization and update for recognized objects.

\textbf{The feature point extraction and tracking pipeline,} basis of visual tracker, starts processing the camera frames upon launching of the mobile client. A sufficient amount of feature points are extracted by the algorithm proposed in \cite{shi1994good} at the beginning, composing a feature points set, and the coordinates of these feature points in following frames are updated via optical flow. 

To support multi-object tracking, as shown in Figure \ref{fig:multitrack}, an array called ``bitmap" is also generated at the beginning, whose length is exactly the same as the number of feature points. This array is used to record the existence and group belonging of each feature point, where for each array element, ``0" means the corresponding feature point is lost during the tracking process, and other positive value (e.g., object ID) indicates the corresponding feature point drops inside the contour of a specific object. After feature point extraction in the first frame, all bitmap values are initialized to 1 to indicate the existence of these feature points, but the group belongings of them are uncertain yet. 

In visual tracker, a group of feature points are extracted and tracked, but it is a rare case for visual tracker to use the same set of feature points from the beginning to the end. Feature points regeneration may happen under two conditions. First, when new objects are recognized by the server. This indicates either new objects have moved into the camera view, or the camera view has changed dramatically. Second, when the number of feature points decreases to a certain threshold. This is caused by lost of some feature points during tracking, or by camera view changing as well.

When a camera frame meets any of these conditions, a new set of feature points are extracted from this frame. The benefit of this design is that for any pair of consecutive frames, there are always enough number of corresponding feature points from same feature points set, which enables uninterrupted object pose estimation.

\textbf{The object pose initialization and update pipeline}, working on top of the former pipeline, starts from the first offloading result. Considering any pair of frames, if we know the old pose of an object and the old locations of feature points within its contour, we can calculate the new pose of the object with the new locations of these feature points using geometric transformation. 

As shown in Figure \ref{fig:frames}, 30 camera frames compose one logical cycle of this pipeline. Based on their functionalities, camera frames can be divided into three categories: \textit{key frames}, \textit{continuous frames} and \textit{result frames}. The first frame of each cycle is a key frame, where a recognition request is sent. At the same time, visual tracker keeps a copy of the current feature points. After several continuous frames, the recognition result is received, and that frame becomes a result frame. The preserved feature points from key frame, together with feature points from result frame, are used to calculate the new poses of the recognized objects within result frame. 

In visual tracker, target objects would go through three states: a) position initialization, as described above; b) position tracking, via geometric transformation between consecutive continuous frames; c) position update, with new recognition results from latter result frames. 
We design the logical cycle for the two reasons. First, new target objects may appear continuously within view. Apparently, cold start latency for recognizing new objects can be shortened by increasing the offloading frequency, but the logical cycle time should be longer than offloading latency. We choose 30 frames ($\sim$ 1 second) as the cycle length considering the offloading latency under different networks.
Second, drift problem should always be taken into consideration during the tracking process, making augmented contents move away from right position gradually. With regular recognition requests, positions of objects can be corrected with recent recognition result, which is more accurate.

With the cooperation of this two pipelines, offloading delay between the key frame and the result frame is hidden from the perception of users, making our proposed tracking method especially suitable for MAR applications under cloud offloading scenario.

\subsection{Content Renderer}
\label{sec:contentRenderer}
Content renderer is the output module of CloudAR. A 3D graphics engine is integrated to render the virtual contents with 6DoF poses of objects, so the virtual contents would attach precisely to physical objects in a 3D world space. For each camera frame, the graphics engine updates the virtual contents based on the output of visual tracker, where for each object a homography is first calculated to find the 2D transition between the reference image and the target image, and the 2D position is later casted into a 3D spatial pose.

\subsection{Network Module}

Network module is the messager between client and server. UDP, as a connection-less protocol, saving the overhead of handshaking dialogues, becomes our choice. Packet losses may happen as UDP is an unreliable communication method, but our system tolerates packet loss naturally: cloud recognitions happen periodically, and new recognition result can be received very soon in next cycle. Even if a reliable network protocol is used for visual task requests, the results could come late due to the re-transmission of lost packets, which is not acceptable in this real time system. 
CloudAR uses non-blocking UDP in independent threads so that the client and server will not be blocked for network issues.

\section{Server design}
\label{sec:serverDesign}
In mobile AR systems, continuous visual processing tasks is nearly impossible in real-time on current devices. In CloudAR framework, we utilize cloud and edge servers as a processing center for those heavy computer vision tasks, which enables the mobile and wearable clients to provide fluent AR experience and abundant contents. 

\subsection{Task Manager}
\label{sec:taskManager}
Upon launching an AR application, the mobile client will communicate with the server to initialize the cloud service. On server side, task manager creates three threads for processing continuous requests from each incoming client: one requests receiving thread, one visual processing thread and one results sending thread. 

\subsection{Visual Task Processor}
\label{sec:visualtaskprocessor}

\begin{figure}[t!]
\center 
\begin{subfigure}[t]{0.23\textwidth}
    \includegraphics[width=\textwidth]{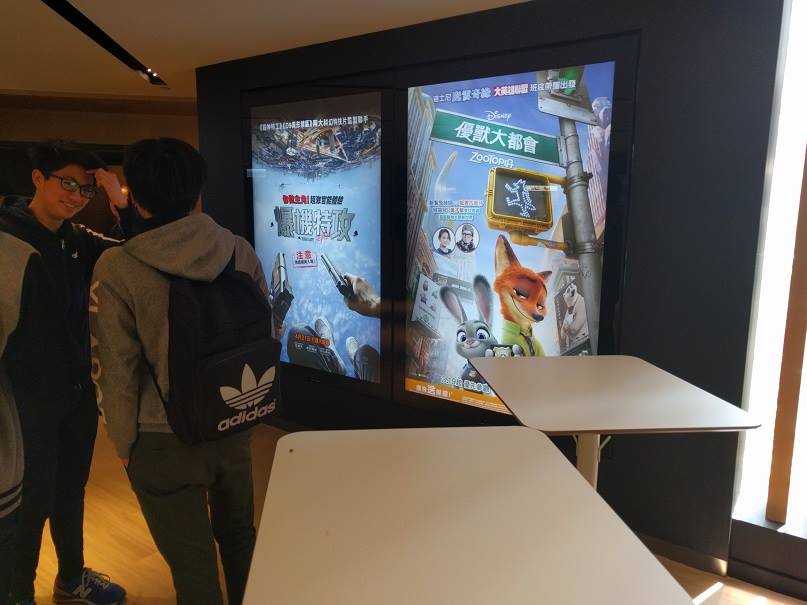}
    \caption{Camera frame input sent from the client to the server.}
    \label{fig:segmentation1}
    \end{subfigure}
\begin{subfigure}[t]{0.23\textwidth}
    \includegraphics[width=\textwidth]{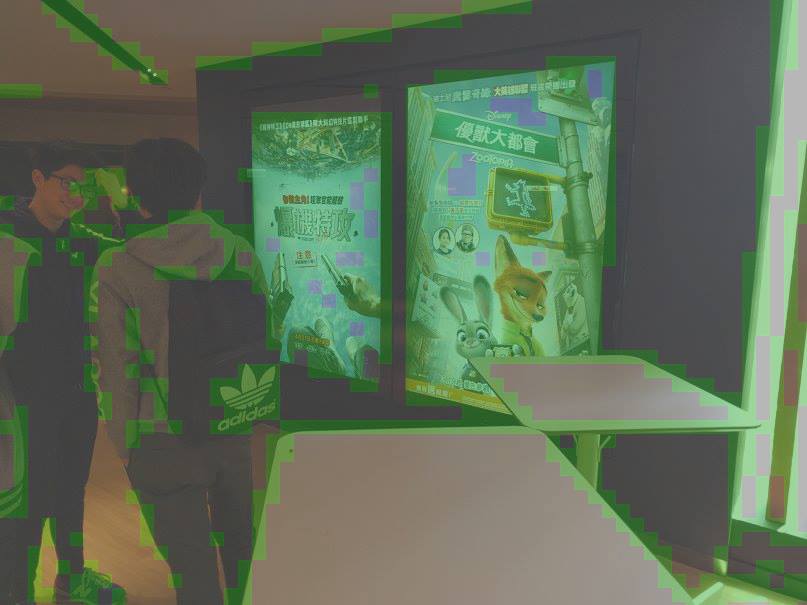}
    \caption{Areas with high pixel variance are highlighted.}
    \label{fig:segmentation2}
    \end{subfigure}
\begin{subfigure}[t]{0.23\textwidth}
    \includegraphics[width=\textwidth]{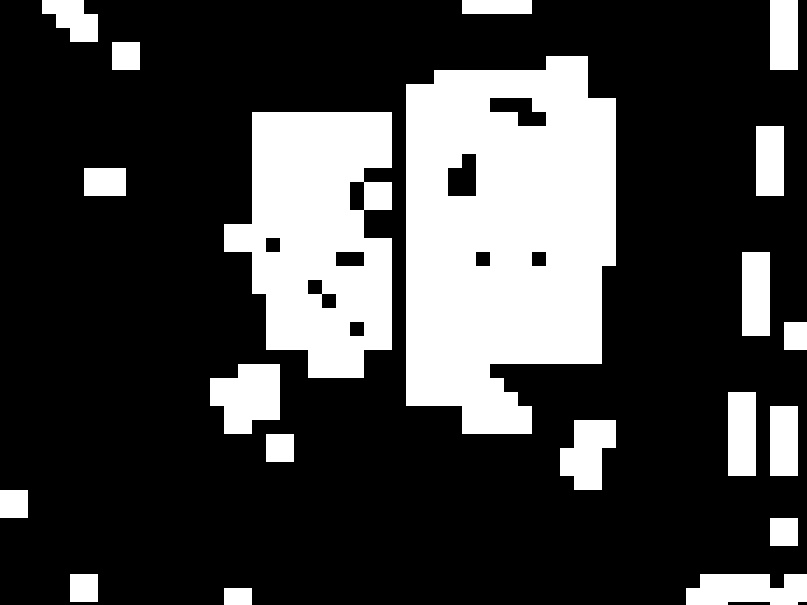}
    \caption{Identified areas before noise reduction as binary image.}
    \label{fig:segmentation3}
    \end{subfigure}
\begin{subfigure}[t]{0.23\textwidth}
    \includegraphics[width=\textwidth]{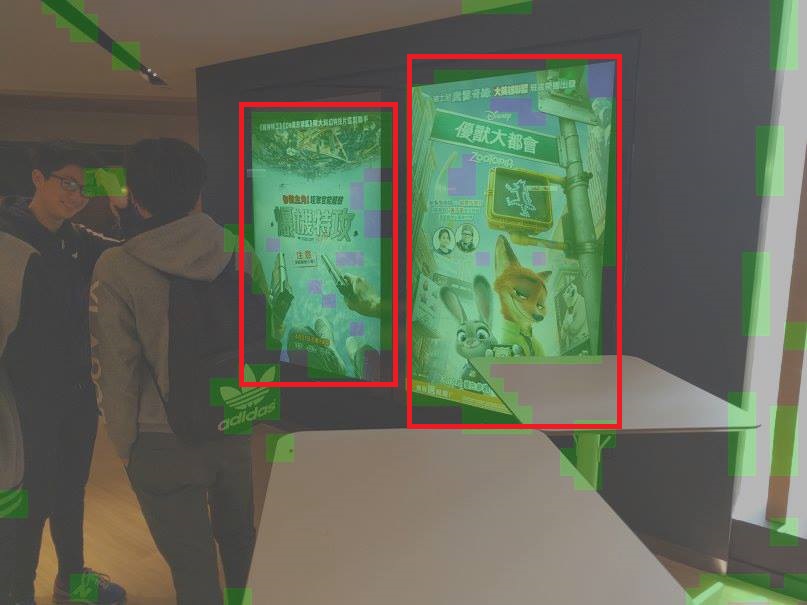}
    \caption{The objects of interest are identified and segmented.}
    \label{fig:segmentation3}
    \end{subfigure}
  \caption{Procedures of the image segmentation.}
  \label{fig:image_segmentation}
\end{figure}

Visual task processor fulfills visual requests, extracting desired information out of camera frames. In our approach, image features are used for retrieving corresponding image from the dataset.

\subsubsection{Image Segmentation}

Upon receiving an incoming camera frame from the client, the server will proceed to perform the image processing tasks. To reduce system load and speed up the process, image segmentation is carried out on the camera frame before retrieving information from it. The image segmentation identifies and extracts the segments in images that contain the image objects, while removing the background in the process. This effectively ensures each query patch contains at most one reference. 

The segmentation process is shown in Figure \ref{fig:image_segmentation}. 
We first apply Gaussian blur to the raw image to reduce noise. Then we apply a sliding window that iterates through the image, and compute the variance in pixel color values for each window. The areas with variance higher than a constant threshold are flagged as positive, while the areas with low variance are flagged as negative since they are likely to be part of the background. This results in a binary image, and we further clean it up by applying an erosion followed by a dilation. The processed image would contain polygons outlining the located objects. The system can then extract the images of the objects by cutting the polygons from the raw image.

\subsubsection{Image Retrieval}

\begin{figure}[t]
    \centering
    \includegraphics[width=0.5\textwidth]{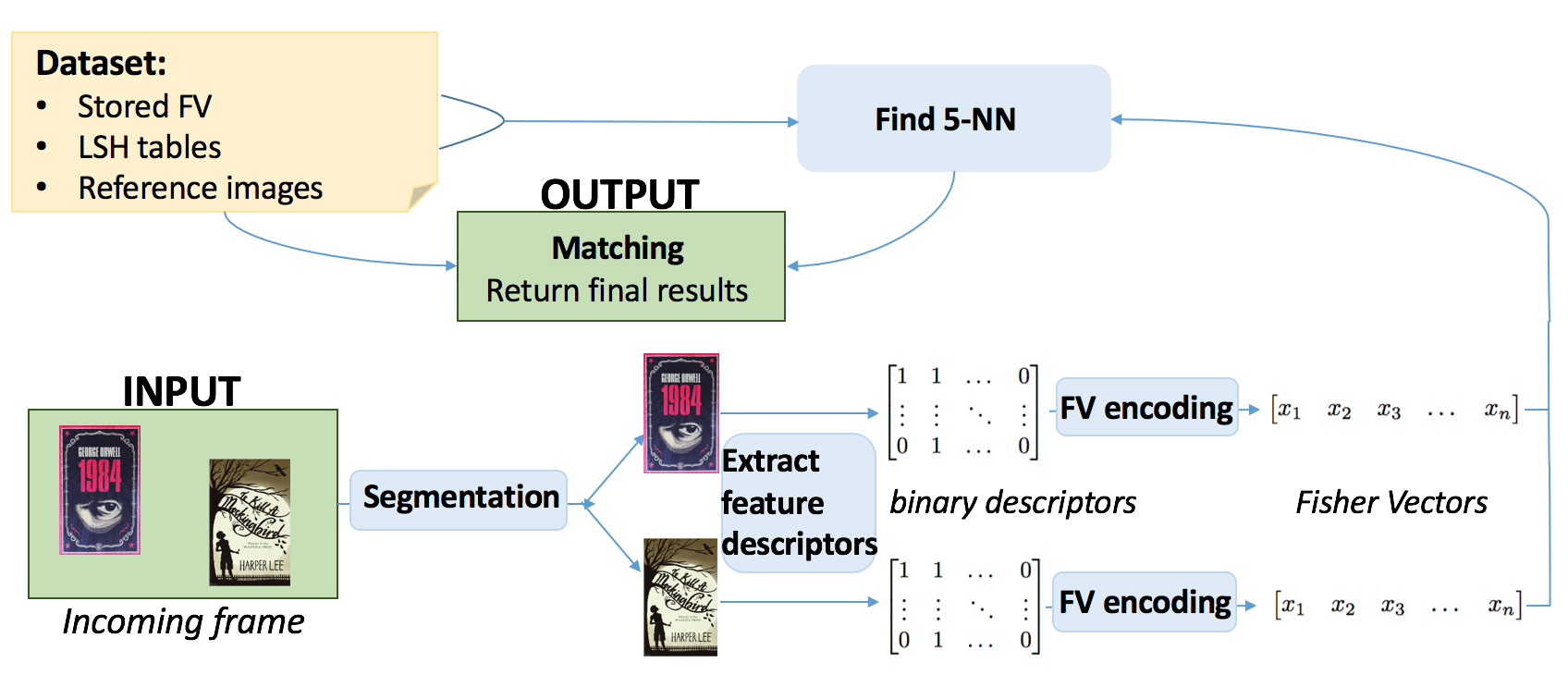}
    \caption{Image recognition pipeline. One frame is first segmented into multiple images, on which image retrieval techniques are applied to find nearest neighbors. Finally, template matching with neighbors verifies the results and calculates positions of the segmented images. }
    \label{fig:recognition}
\end{figure}

The overall image retrieval pipeline is shown in Figure \ref{fig:recognition}. To build the dataset, we need to detect key points and extract corresponding features from each reference image. As the reference images for building the dataset and the query images getting from camera frames need to be handled through the same procedure, we need to consider the processing speed. Multi-scale AGAST\cite{mair2010adaptive} detector and FREAK\cite{alahi2012freak} binary feature descriptors are used, and we adopt Fisher encoding built upon binary features for accurate and real-time performance. To do Fisher encoding on binary features, we first need to build a Bernoulli mixture model (BMM), 
 which is built upon the collection 
 of all descriptors 
 of the dataset images. 
After getting parameters of the BMM model, descriptors of an image 
can be encoded into a single Fisher Vector (FV). Encoded FVs will be L2 and power normalized, and all FVs of the dataset images will be stored and go through Locality Sensitive Hashing (LSH) to create hash tables for faster retrieval.

After building up the dataset, server can take the responsibility for recognition tasks. 
Segmented individual patches will be simultaneously re-sized into similar scale to the reference images, and encoded into FVs through the procedures mentioned above. The parameters for the BMM model used for query patch Fisher encoding is the same as that used for encoding reference images, which have been already calculated. By LSH tables and stored FVs, we will find the nearest neighbors of the segmented patches. Five nearest neighbors will be found for each patch, and feature matching would be used to select the correct one from the five. 

Feature matching is the last step to verify the result of image retrieval and calculate a 6DOF pose of the target. For each segmented patch, with the feature descriptors from both the patch and the nearest neighbors, feature matcher can find the corresponding feature pairs. If enough number of matches are found for any neighbor, we pick that neighbor as the recognition result and start pose estimation. We calculate the homography of the target within camera view, which stands for a 2D position transition, and return only simple corner positions to the task manager for sending it back to the mobile client. If there are not enough good matches for all five nearest neighbors in the dataset, the patch will be discarded and no result will be returned.

\section{Implementation and Evaluation}
\label{sec:evaluation}
We implement the mobile client on the \textit{Android} platform. The mobile visual tracker is implemented with \textit{OpenCV4Android}\footnote{\url{http://opencv.org/platforms/android.html}} library, and the 3D renderer is a modified \textit{Rajawali}\footnote{\url{https://github.com/Rajawali/Rajawali}} library. 
The cloud server side is implemented in C++ based on Linux platform. \textit{OpenCV}\footnote{\url{http://opencv.org}} library is used for multi-scale AGAST feature points detection and FREAK descriptor extraction. \textit{FALCONN}\footnote{\url{https://falconn-lib.org}} library is used for building LSH tables and finding approximate nearest neighbor of FVs. We also rely on \textit{OpenMP}\footnote{\url{http://openmp.org}} for parallel processing.

\begin{figure}[t]
    \centering
    \includegraphics[width=0.5\textwidth]{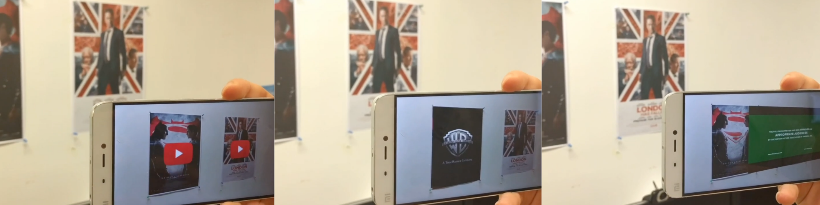}
    \caption{The PosterAR App running our CloudAR framework. The App utilizes our framework to retrieve poster information and overlay movie trailers on the camera stream.}
    \label{fig:run}
\end{figure}

We develop an AR application called PosterAR using our CloudAR framework. When a user steps into a cinema, he/she does not have a clear idea about which movie to watch. At this time the user can launch the PosterAR App, point the camera to the posters, and watch a trailer of that movie. The play buttons and movie trailers are strictly overlaid according to the homography of the posters, so that they are precisely coupled as if there is a virtual video wall.

The running examples of Poster App are shown in Figure~\ref{fig:run}. On the left the user is holding the phone against two movie posters and the App augments trailers on top of the posters. On the middle and right, movie trailers are played in 3D world space once the user presses the play button. In this section, we collect relevant data from PosterAR to evaluate the performance of our framework.

\begin{figure*}[t]
    \centering
 \includegraphics[width=1\linewidth]{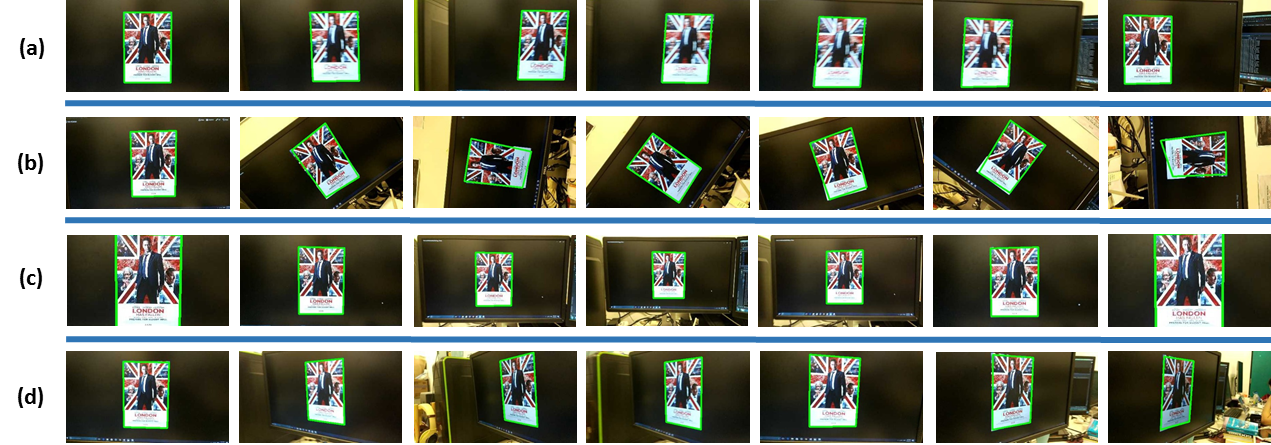}
    \caption{Four common scenarios of camera movement: (a) fast movement with motion blur; (b) camera rotate; (c) scaling caused by forward and backward movement; (d) tilt caused by changing view. The green edges are labeled with tracking results from the mobile client.}
    \label{fig:films}
\end{figure*}

\subsection{Experiment Setup}
\label{sec:experimentSetup}
On the client side, the PosterAR App runs on a Xiaomi MI5 with a quad core CPU and 4GB RAM. On the server side, we deploy servers on both local PC and Google Cloud Platform, considering the requirements of edge computing and cloud computing. The local PC is configured with an Intel i7-5820k CPU (6 cores @3.3GHz) and 32GB RAM, and we create a WiFi Access Point on the PC so that the phone is directly connected to the local server. This setup reflects the scenario of edge computing as the processing components are placed nearby in the LTE tower or local router within one hop distance of the mobile device. Meanwhile, the virtual machine on Google Cloud is configured with 8 vCPUs and 32GB RAM. 

We conduct our experiments with respect to mobile tracking performance, offloading latency, and image retrieval accuracy.

\subsection{Tracking Performance}
\label{sec:trackingPerformance}
Visual tracker is the key component of the CloudAR framework, whose performance affects directly the user's perceived experience. In this section, we evaluate the performance of visual tracker with respect to the tracking accuracy and tracking speed.

\subsubsection{Tracking Accuracy}
An important indicator of the tracking performance is the quality of 6DoF tracking, which is crucial in providing a seamless AR experience. We design four common tracking scenarios to evaluate the performance: \textit{fast movement}, \textit{rotation}, \textit{scaling}, and \textit{tilt}. The results are shown in Figure \ref{fig:films}.

\begin{figure}[t]
\center 
\begin{subfigure}[b]{0.24\textwidth}
    \includegraphics[width=\textwidth]{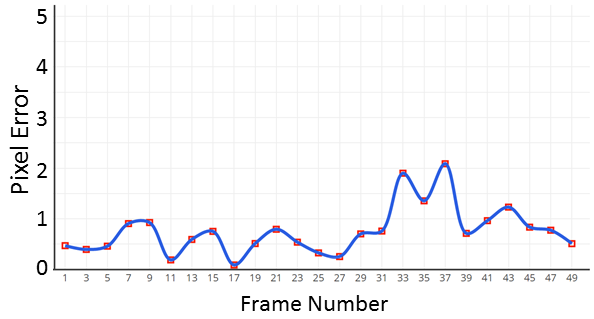}
    \caption{Fast Movement - Pixel errors.}
    \label{fig:moveres}
    \end{subfigure}
\begin{subfigure}[b]{0.24\textwidth}
    \includegraphics[width=\textwidth]{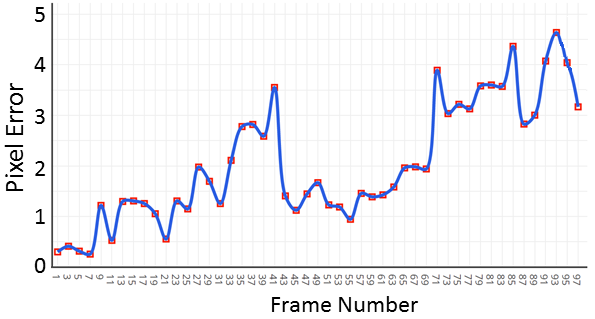}
    \caption{Rotation - Pixel errors. }
    \label{fig:rotateres}
    \end{subfigure}
\begin{subfigure}[b]{0.24\textwidth}
    \includegraphics[width=\textwidth]{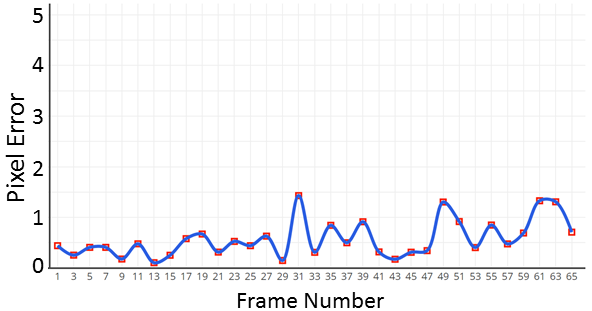}
    \caption{Scaling - Pixel errors.}
    \label{fig:scaleres}
    \end{subfigure}    
\begin{subfigure}[b]{0.24\textwidth}
    \includegraphics[width=\textwidth]{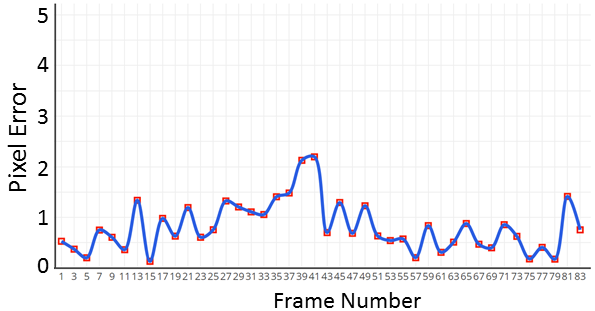}
    \caption{Tilt - Pixel errors.}
    \label{fig:tiltres}
    \end{subfigure}    
   \caption{Tracking accuracy in pixel errors. Our visual tracker performs good in all four scenarios of camera movement with little pixel errors. }
  \label{fig:trackres}
\end{figure}

We record the positions of the poster's four corners in each frame given by the tracker, and obtain the ground truth by manually labeling corners for all frames. There are two common ways to measure the tracking accuracy: a) \textit{intersection over union (IOU)}, which considers the bounding boxes of both the tracking result and the ground truth, and it is defined as the proportion of intersection area within the union area of the two boxes; b) \textit{pixel error}, which is defined as the pixel distance on screen between the center of the tracking result and that of the ground truth. As our tracking method provides fine-grained 6DoF results, the bounding boxes are irregular quadrilaterals, whose intersection area and union area are complex to calculate. Instead, the center of each bounding box is more convenient to find, which is the intersection point of the two diagonals of the box. For this reason, we use \textit{pixel error} as the evaluation metrics in this part. The results are shown in Figure \ref{fig:trackres}, where the pixel errors in all scenarios are as low as 1 or 2 for most of the time. The worst case is rotation, but the pixel errors are still under 5. 

Also the rotation scenario shows a trend of drift, which is common in incremental tracking. However, our system handles the drift problem as the mobile client sends recognition requests periodically, and revises the position in tracking with the recognition results accordingly. This design guarantees that drift in tracking will last no longer than 30 frames in the worst case. 

\subsubsection{Tracking Speed}

The time consumption of tracking each frame is composed of four parts: time to down sample the frame, time to calculate the optical flow of feature points, time to estimate the poses of objects, and idle time waiting for next data. Since we use feature points based method in tracking, the amount of points would influence the performance and delay of tracker. To learn the impact of this factor, we measure the time consumptions of tracking camera frames with four different numbers of feature points, which are 60, 120, 180 and 240, correspondingly. For each case, we record the timestamps of different parts for 500 frames. 

\begin{figure}[t]
    \centering
    \includegraphics[width=0.3\textwidth]{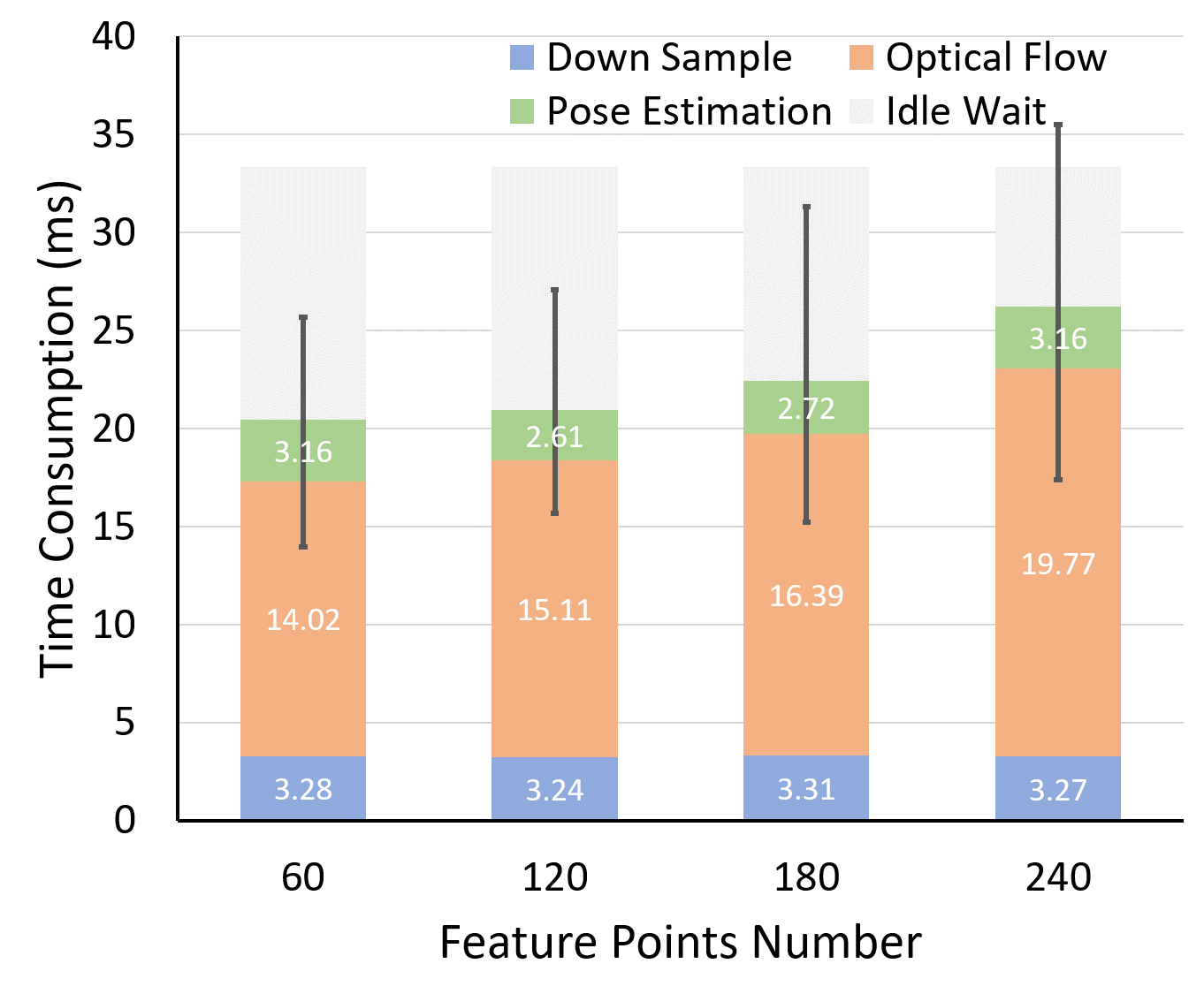}
    \caption{Time consumption for tracking one frame with different amount of feature points. The results are averaged over 200 runs with standard derivation shown in black line.}
    \label{fig:tracktime}
\end{figure}

The results are shown in Figure~\ref{fig:tracktime}. There is a clear trend that the time consumption of tracking one frame increases with the number of feature points, and the optical flow calculation part results in this time variance. To guarantee a 30 FPS performance of the tracker, the upper limit of the time consumption on each frame is 33.34ms. We can see that all four cases satisfy this requirement on average, and non-qualified only in worst cases with 240 feature points. We conclude that our visual tracker ensures a real-time performance with the amount of feature points less than or equal to 180.

\subsection{Offloading Latency}
\label{sec:latency}

Offloading latency is defined as the time period from the moment obtaining request camera frame data to the moment displaying offloading result, which is composed of three parts: delay of client, including the time to compose request and calculate new poses after receiving the result; delay of server, including the time to parse request and recognize image; delay of network, including the round trip time of the datagram packets.

In this section, we measure both the local offloading latency and the cloud offloading latency with the PosterAR App. For each offloading scenario, 200 offloading tasks are generated and recorded.

\subsubsection{Latency Distribution}

\begin{figure}[t]
\center 
\begin{subfigure}[t]{0.24\textwidth}
    \includegraphics[width=\textwidth]{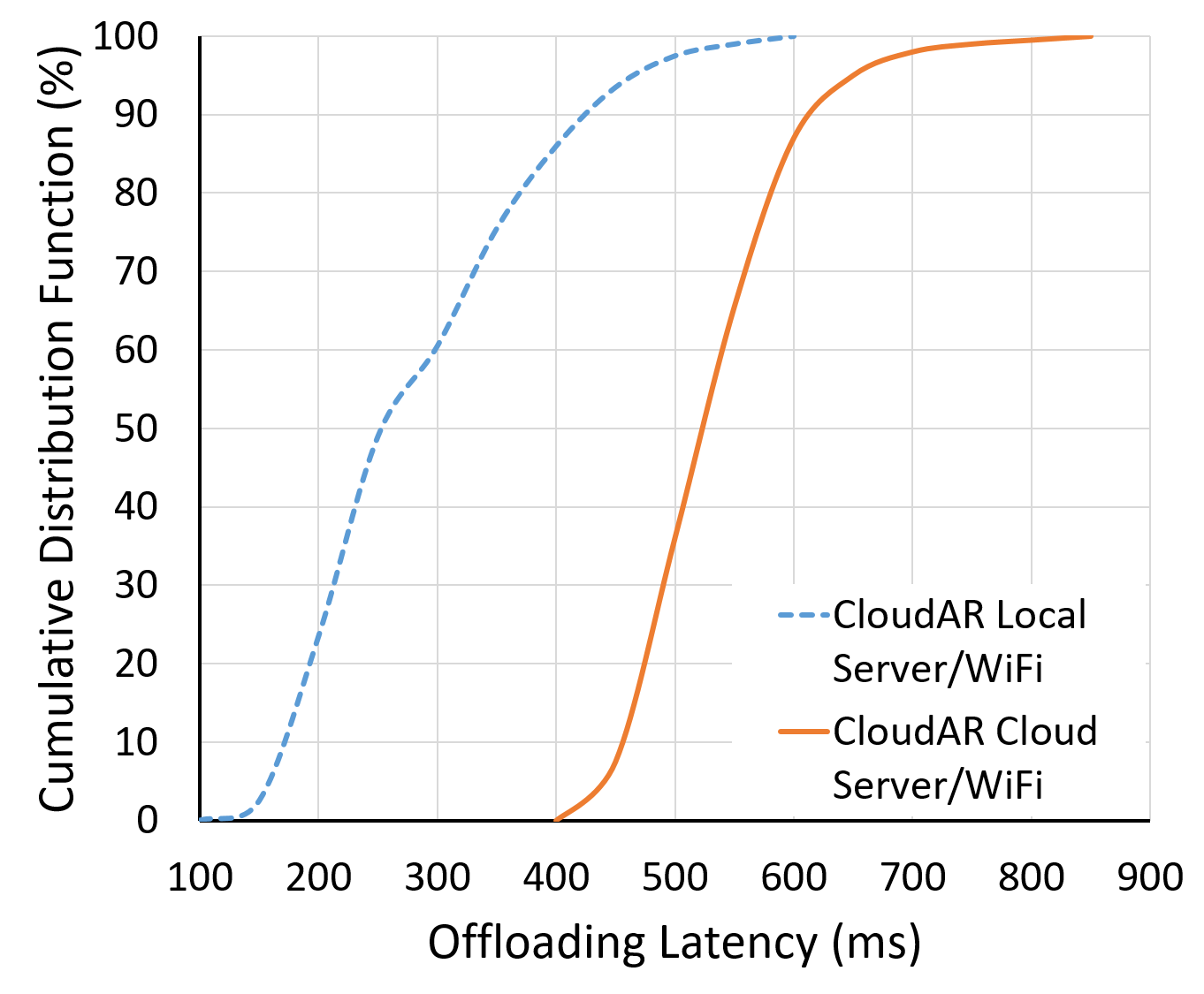}
    \caption{CDF of offloading latency for both local offloading scenario and cloud offloading scenario.}
    \label{fig:offloadcdf}
    \end{subfigure}
\begin{subfigure}[t]{0.24\textwidth}
    \includegraphics[width=\textwidth]{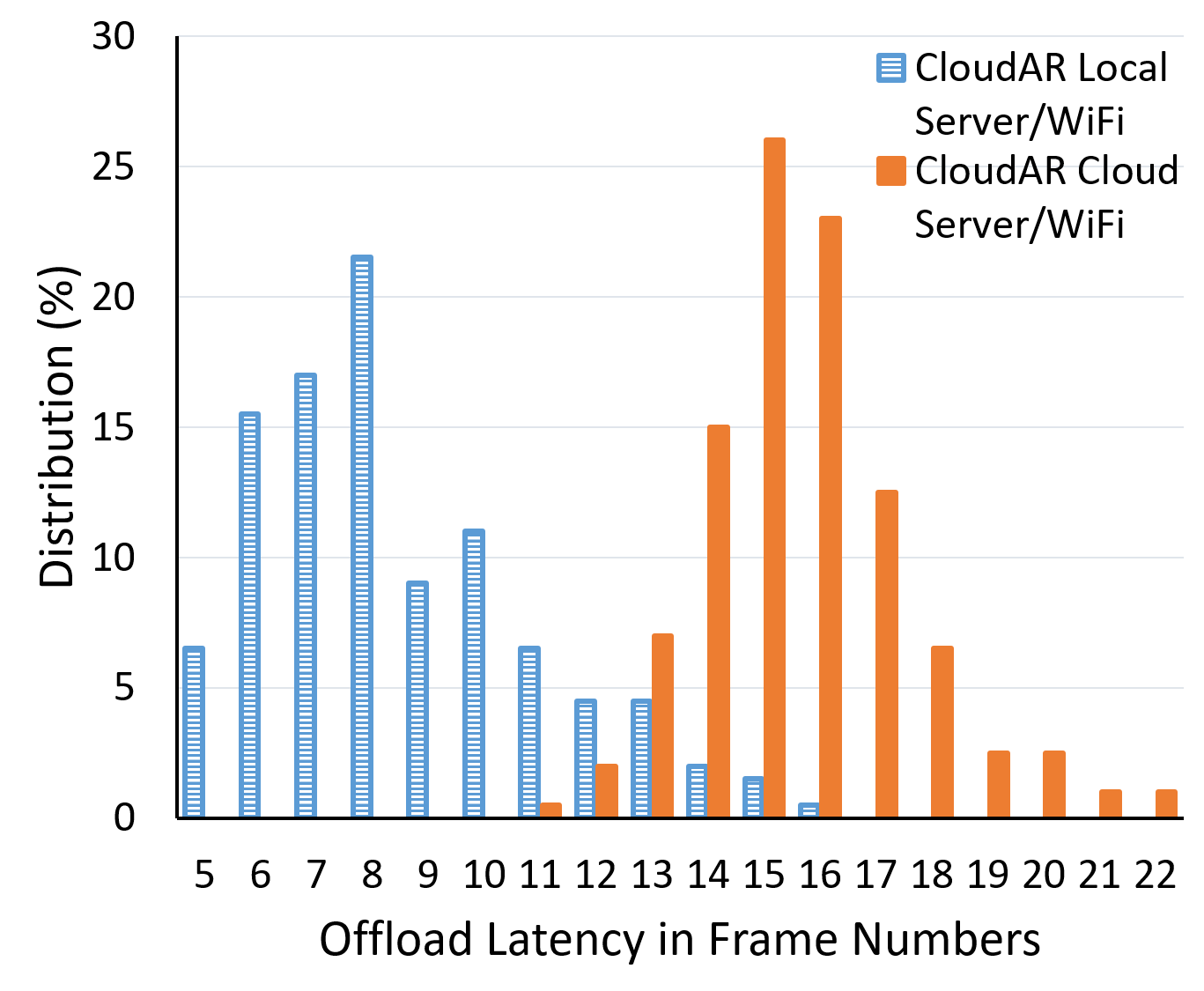}
    \caption{Distribution of offloading latency in passed frames. }
    \label{fig:offloadframe}
    \end{subfigure}
  \caption{Delay of offloading for both local scenario and cloud scenario with 200 runs. Compared with cloud offloading scenario, local offloading scenario tends to have a lower latency, as shown in both latency time and passed frames.}
  \label{fig:delay}
\end{figure}

Figure~\ref{fig:offloadcdf} is the CDF of overall offloading latency for two offloading scenarios, and local offloading scenario tends to have a lower latency than cloud offloading scenario. 60\% of the local offloading tasks are finished before 300ms, and 97\% of them are finished before 500ms. On the contrary, only 35\% of the cloud offloading tasks are finished before 500ms, but all of them are finished before 850ms. (We also measure the latency of cloud offloading scenario under 4G connection in Section \ref{sec:cloudlatency}.) Figure~\ref{fig:offloadframe} is the distribution of offloading latency in passed frame numbers, which represents the number of frames that has passed (processed by visual tracker) before the offloading results are showed to user. For local offloading scenario, 60\% of the results are showed within 8 frames after the starting points of offloading, and 100\% of them are showed within 16 frames. Meanwhile, the distribution of cloud offloading scenario looks like a normal distribution, and we come to the same conclusion as before that cloud offloading scenario has a longer latency than local offloading scenario, as the mean of this ``normal distribution'' is 15 frames.

\begin{figure}[!tb]
    \centering
    \includegraphics[width=0.5\textwidth]{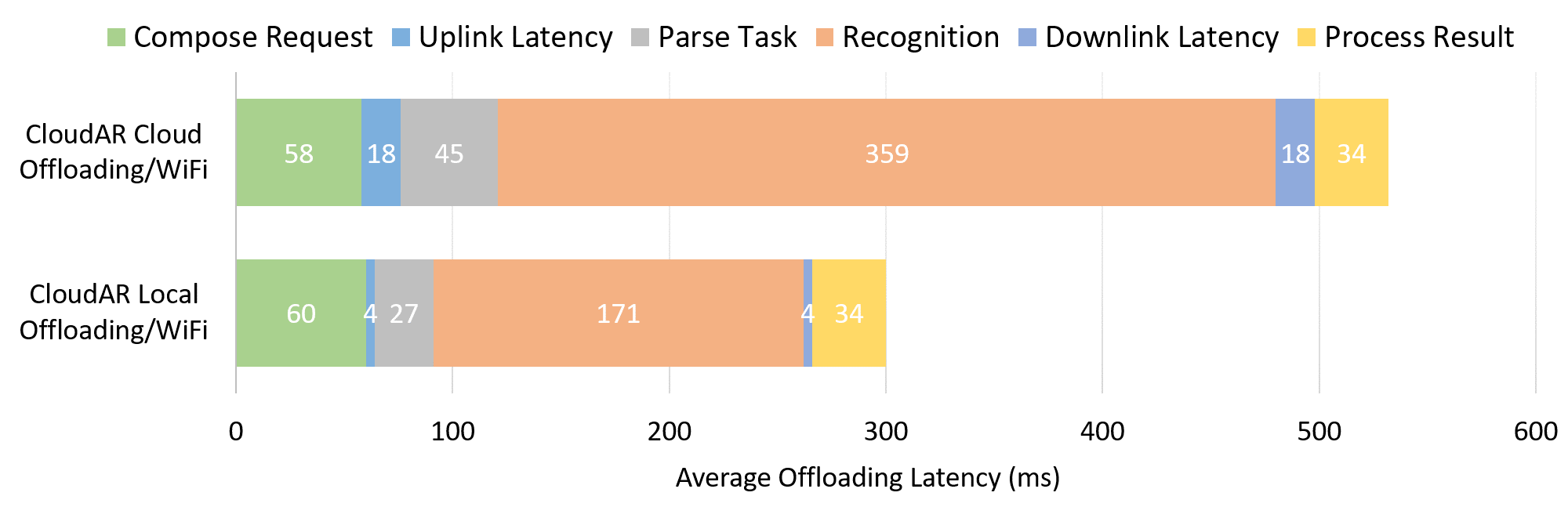}
    \caption{Components of latency for both local offloading scenario and cloud offloading scenario. The results are averaged over 200 runs.}
    \label{fig:offloadlatency}
\end{figure}

Figure \ref{fig:offloadlatency} shows the components of latency for both scenarios. The part \textit{compose request} takes around 60ms, where encoding raw frame data into a small image file spends most of this time. The \textit{uplink latency} and \textit{downlink latency} parts in local offloading scenario are much shorter than that in cloud offloading scenario. \textit{Parse task} is the first part of time spent on server to parse the request as well as control the processing. \textit{Recognition} is the second part of time spent on server, including the time to segment the camera frame, find nearest neighbor, match and verify the result. For both \textit{parse task} and \textit{recognition}, the local server processes faster than the cloud server, showing that the local server has a better processing capability. \textit{Process result} is the time that the client processes the result and displays it to user.

From the requirements of visual tracker, the whole offloading procedure should finish within 30 frames (1 second). The results from figure \ref{fig:offloadlatency} prove that our offloading pipeline fulfills this requirement under both local and cloud offloading scenarios.

\subsection{Image Retrieval Accuracy}
\label{sec:imageRetreivalAccuracy}

\begin{figure}[t]
    \centering
 \includegraphics[width=0.3\textwidth]{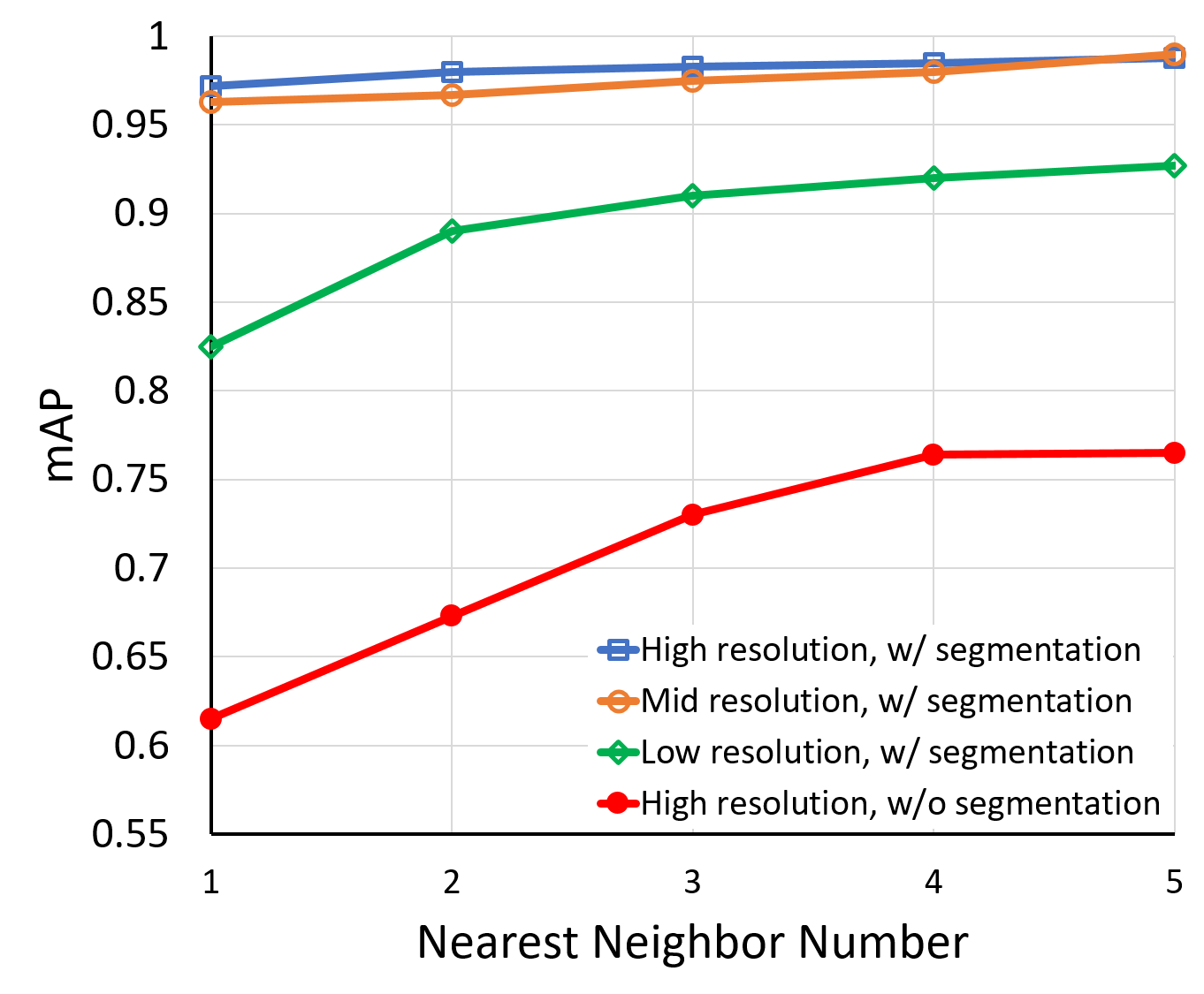}
    \caption{Image retrieval mean Average Precision (mAP) for matching 1-5 nearest neighbors (NN) with or without segmentation under different scales.}
    \label{fig:map}
\end{figure}

We collect several hundreds of movie posters online, combining with some data from Stanford Mobile Visual Search Dataset \cite{chandrasekhar2011stanford} to form 1000 reference images. Query sets are photoed by Mi5's back camera, together with several query sets from \cite{chandrasekhar2011stanford}.

If the query images are passed directly into the feature extraction stage without segmentation, then the resulting retrieval mean Average Precision (mAP) is around 0.76 for top-5 NN (nearest neighbor), and around 0.61 if we only check one most nearest neighbor, as shown by the ``without segmentation'' line in Figure~\ref{fig:map}. The relative low mAP is mainly caused by multi-target existence within one query image, or random scattered noise in the background.

Given good segmentation results, the retrieval can achieve good accuracy for top-5 NN. In our experiment, all patches will be re-sized into a 400$\times$400 resolution. From Figure \ref{fig:map} we can see that in most cases, we only need to examine the most nearest neighbor during the matching step to get the correct result, and the expectation of examined neighbor number is close to 1. The high-resolution line corresponds to a patch size of at least 400$\times$400 pixels. The mid-resolution line corresponds to having segmented patches with size around 200$\times$200 pixels, and the small-resolution line corresponds to patch size of 100$\times$100 pixels. During our experiment, more than 90\% segmented incoming frames can correctly retrieve their corresponding images in the dataset.

\subsection{Performance Comparison with Vuforia}
\label{sec:overallPerformance}

\textit{Vuforia} is a leading commercial AR framework that supports cloud recognition, and CloudAR focuses on cloud-based AR solutions. Therefore, we compare several performance metrics of cloud recognition functionalities between CloudAR and Vuforia. 

To setup the experiment, we download the official CloudReco App from Vuforia. This App performs image recognition tasks by offloading, receives results and displays augmented contents to user. 
On the Android side of Vuforia, we do essential instrumentation on CloudReco to log analysis data and influence the performance as little as possible. On the server side of Vuforia, we create a image database with their web services. Since Vuforia supports only one-by-one image uploading, we create a database with only 100 images. For our PosterAR App, we use the previous setup mentioned before, and the cloud server hosted on Google Cloud Platform is used to pursue a fair comparison. Both WiFi and 4G connections of the phone are used to evaluate the performance of the two frameworks under different network conditions.

\subsubsection{Data Usage}
\label{sec:datausage}

Data transmission is always involved in offloading, so we want to compare the data traffic usage for the two frameworks. Through our network monitoring of the CloudReco App, we find out that cloud recognitions in Vuforia framework are implemented with Https requests. With this knowledge, we setup a man-in-the-middle (MITM) proxy using Fiddler8 and decrypt all the recognition requests and responses of the CloudReco App. Our key finding is that each request body contains a JPEG file of the
current camera frame, and each response body contains another JPEG file of the recognized original image from database. Since there are only 100 images in our created Vuforia demo database, we use this 100 images to trigger cloud recognition tasks on both frameworks. The average request sizes of the PosterAR App and the CloudReco App are 11.58KB and 16.24KB correspondingly, as both of them are sending compressed camera frame to the cloud. However, the average result sizes of the two Apps differ hugely. In CloudAR, the results contain only coordinates of the bounding
boxes and other identity information of the images, so we make
the result a fixed size of only 400 bytes. In Vuforia, the average size
of offloading results is 33.48KB, showing that Vuforia consumes
much more data traffic than CloudAR in each offloading task.

\begin{figure}[t]
\center 
\begin{subfigure}[t]{0.24\textwidth}
    \includegraphics[width=\textwidth]{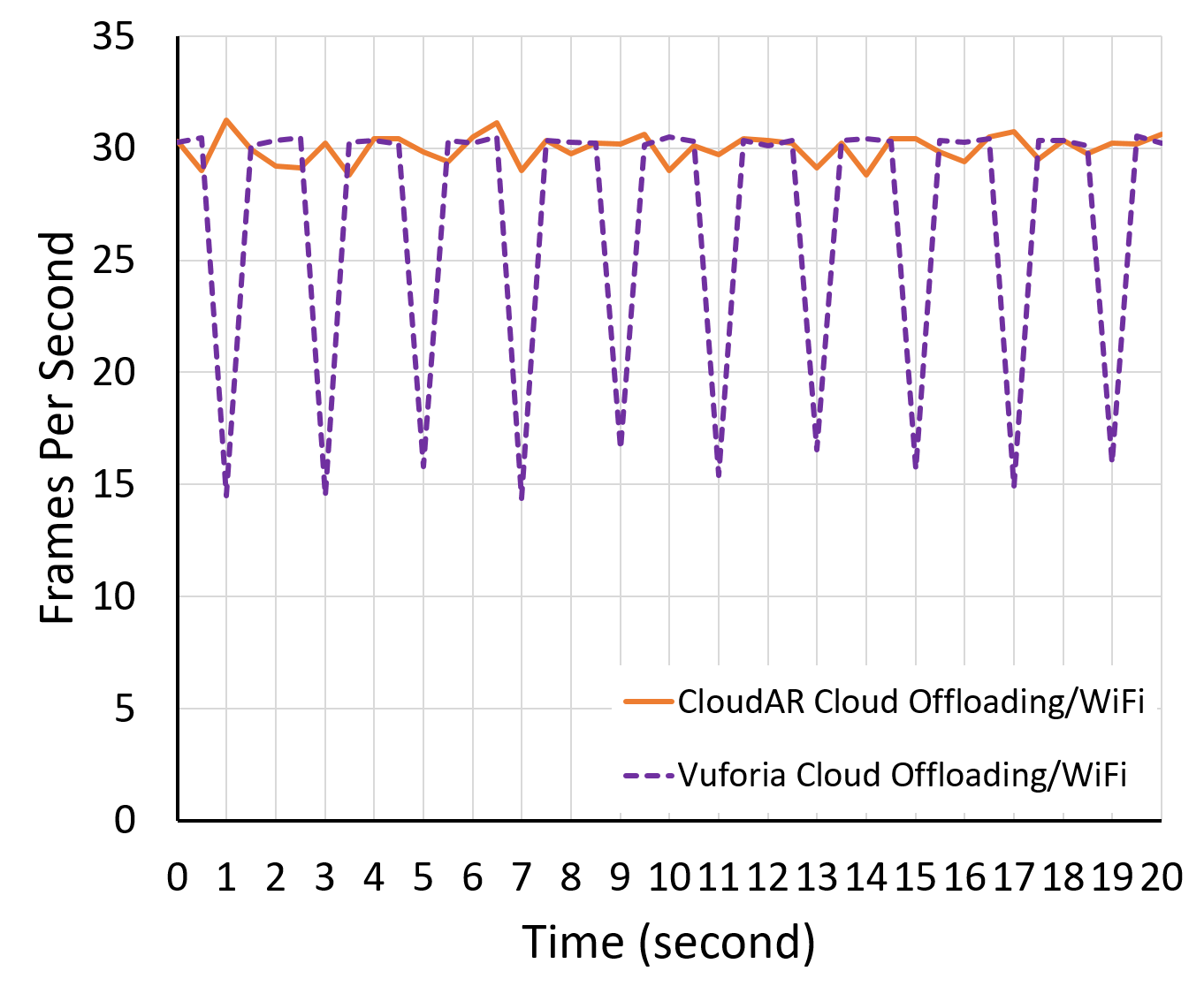}
    \caption{FPS of Apps with offloading built with CloudAR and Vuforia.}
    \label{fig:fps}
    \end{subfigure}
\begin{subfigure}[t]{0.24\textwidth}
    \includegraphics[width=\textwidth]{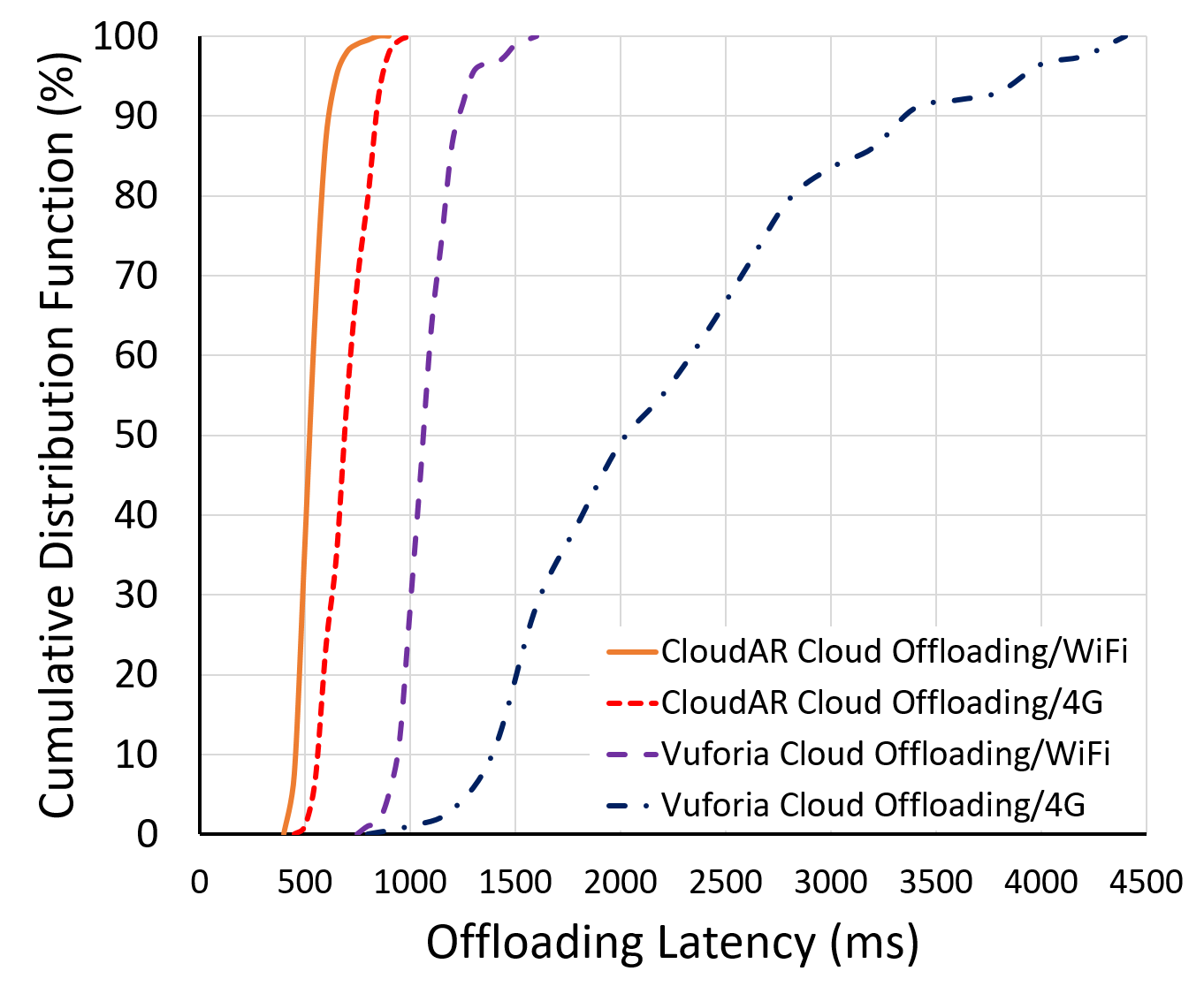}
    \caption{CDF of offloading latency for both CloudAR and Vuforia with 200 runs.}
    \label{fig:vuforialatency}
    \end{subfigure}
  \caption{Performance comparison Of CloudAR and Vuforia in terms of FPS and offloading latency.}
  \label{fig:performance}
\end{figure}

\subsubsection{Runtime FPS}
\label{sec:fps}
FPS reflects the runtime performance of the Apps, which has straightforward influence on the user perceived experience. Offloading procedure is an extra burden onto the  mobile device besides frame-by-frame tracking, so we want to learn how would offloading influence the FPS. 

The offloading procedure of the PosterAR App is periodical, but the CloudReco App would trigger an offloading task when there is no recognized object in the current frame. Imagine a scenario where the user scans through a series of posters on the wall, and periodical offloading happens in both frameworks. To simulate this scenario, we make a slideshow of the images on the screen and point the phone's camera to the screen when the two Apps are launched. The slideshow has a fixed change interval of 2 seconds, and we record the runtime FPS of both Apps. A typical fragment of the result is shown in figure \ref{fig:fps}. We can see that the FPS of the CloudReco App will decrease significantly upon receiving a result, and we perceive obvious non-smooth frames during the experiment. In comparison, the PosterAR App runs at a much stable FPS with little jitters around 30. As the offloading result of Vuforia is the original image, feature extraction and matching are executed on the client, which are computationally intensive for mobile devices that decrease the FPS significantly.  

\subsubsection{Offloading Latency}
\label{sec:cloudlatency}
We use the same scenario in measuring FPS to trigger 200 offloading tasks for both CloudAR and Vuforia under WiFi and 4G connections, and the offloading latency results are shown in figure \ref{fig:vuforialatency}. We have already seen the CDF of CloudAR cloud offloading latency under WiFi connection in previous section, and the CDF of that latency under 4G connection has a similar pattern with it, with slightly longer latency but still guaranteed to finish within 1 second. These latency results of CloudAR cloud offloading prove that our periodical offloading design works with cloud servers under both WiFi and 4G connections. 

On the contrary, Vuforia cloud offloading has a much longer latency, especially under 4G connection. For Vuforia cloud offloading under WiFi connection, only 30 percent of the offloading tasks are finished with 1 second, and the worst case takes around 1.6 seconds. It is worth to mention that there is not much difference in the network accessing latency of Vuforia and CloudAR cloud servers, with RTTs of 48 ms and 36 ms, respectively under WiFi connection. Vuforia cloud offloading under 4G connection has a noticeable poorer performance, with only 50 percent of the offloading tasks are finished within 2 seconds, and the worst case takes as long as 4.4 seconds. This poor performance of Vuforia under 4G possibly comes from the combination of large result size and lossy 4G connection, which turns into a long cold-start latency. 						

\subsubsection{Multi-target Recognition Support}
\label{sec:multitarget}
CloudAR framework supports multi-target recognition from the design of mobile visual tracker and cloud image recognition pipeline, and a running example of recognizing multiple targets is showed in figure \ref{fig:run}. In CloudAR framework, most of the computation burden of recognizing multiple target is offloaded to server, and the mobile client only processes simple tasks like calculating the bounding boxes with offloading results. On the contrary, the CloudReco App does not support multi-target recognition, and only one target within camera view is recognized every time. Even if multi-target cloud recognition is supported under the current Vuforia offloading design, multiple original images would make the results quite huge, and the heavy feature extraction and template matching processes for multiple images on mobile would ruin the user experience.


\section{Conclusion}
\label{sec:conclusion}
In this paper, we presented CloudAR, a cloud-based AR framework which targets at solving the large-scale multiple image recognition and real-time augmenting problems on mobile and wearable devices. We proposed an innovative tracker running on mobile to hide the offloading latency from user's perception, and a multi-object image retrieval pipeline running on server.
Our evaluation showed that the AR App built with CloudAR performs effectively in terms of seamless AR experience and robustness. 
We further showed that our framework has high tracking accuracy, high recognition accuracy, low overall offloading latency, and high runtime performance. 

\section*{Acknowledgment}
  This work was supported in part by the General Research Fund from the Research Grants Council of Hong Kong under Grant 26211515 and in part by the Innovation and Technology Fund from the Hong Kong Innovation and Technology Commission under Grant ITS/369/14FP.


\bibliographystyle{IEEEtran}
\bibliography{main}

%
\IEEEpeerreviewmaketitle

\ifCLASSOPTIONcaptionsoff
  \newpage
\fi



%
%

\begin{IEEEbiography}[{\includegraphics[width=1in,height=1in,clip,keepaspectratio]{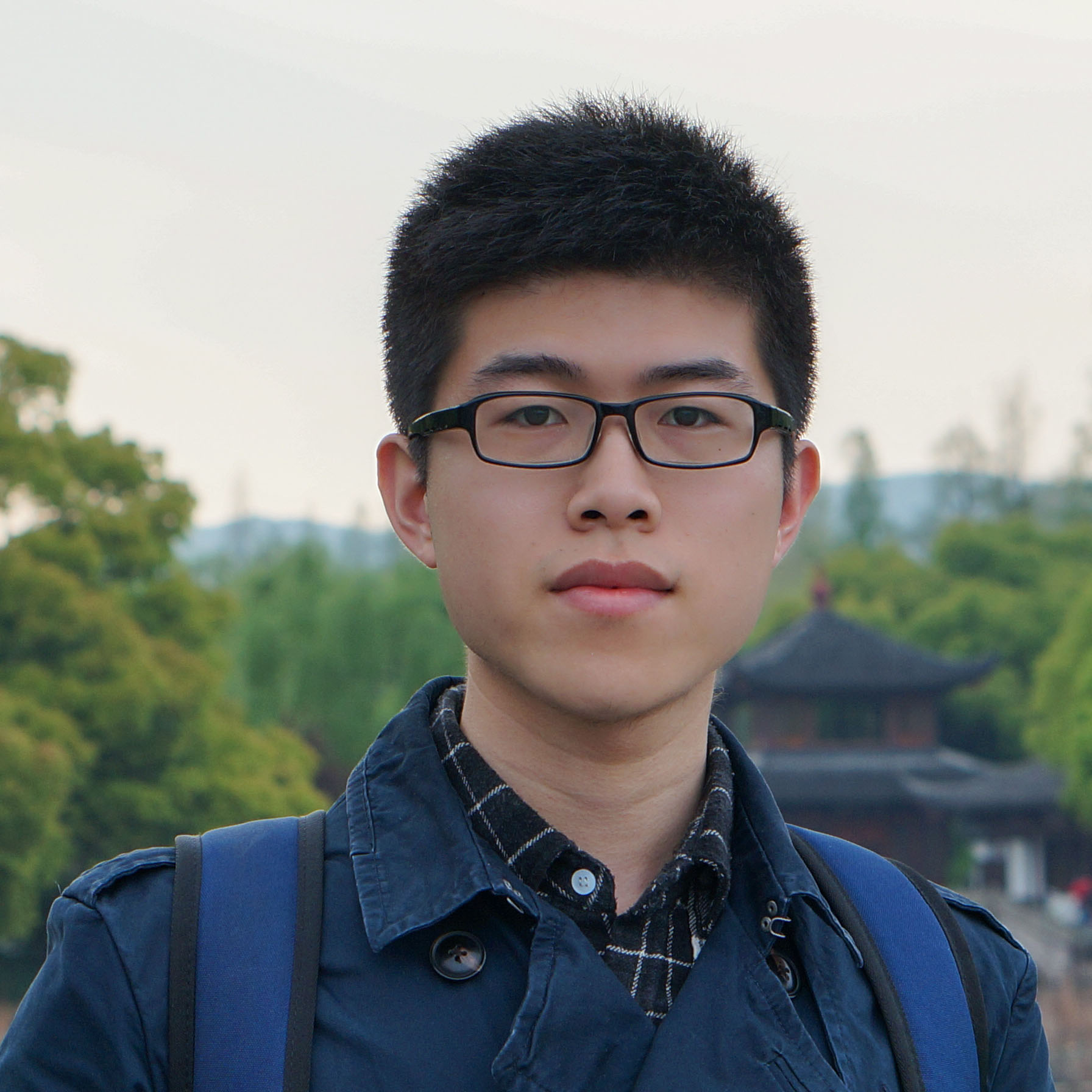}}]{Wenxiao ZHANG}
received the bachelor's degree in computer science
and technology from the Department of Computer Science and Technology at Zhejiang University, China, in 2014. He is currently working toward the PhD degree in the Department of Computer Science and Engineering at The Hong Kong University of Science and Technology and a member of Symlab. His main research interests include the areas of mobile computing, edge computing, and augmented reality.
\end{IEEEbiography}

\begin{IEEEbiography}[{\includegraphics[width=1in,height=1in,clip,keepaspectratio]{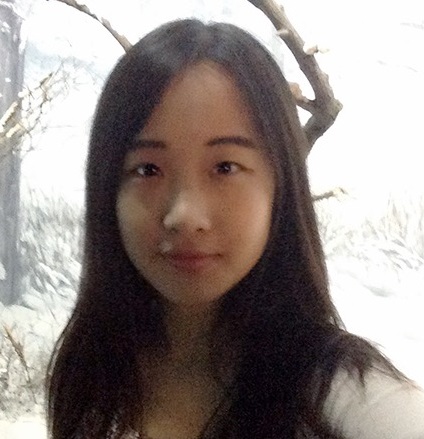}}]{Sikun LIN}
received the bachelor's and master's degrees in computer science and engineering from the Department of Computer Science and Engineering at the Hong Kong University of Science and Technology, Hong Kong, in 2015 and 2017, respectively. She is currently working toward the PhD degree in the Department of Computer Science at University of California, Santa Barbara, U.S.. Her main research interests include the areas of computer vision and human-computer interaction. 
\end{IEEEbiography}

\begin{IEEEbiography}[{\includegraphics[width=1in,height=1in,clip,keepaspectratio]{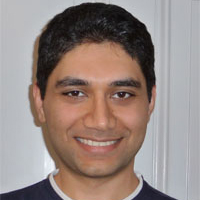}}]{Farshid Hassani Bijarbooneh}
received the master's and PhD degrees in computer science from the Department of Information Technology at the Uppsala University, Sweden, in 2009 and 2015, respectively. In 2015$-$2017, he was a postdoctoral researcher with the Hong Kong University of Science and Technology, Hong Kong. He is currently working as Senior Developer at Assessio and Co-founder of SpaceFox Studio, Stockholm. His main research interests include the areas of mobile computing, wireless sensor networks, and augmented reality.
\end{IEEEbiography}

\begin{IEEEbiography}[{\includegraphics[width=1in,height=1in,clip,keepaspectratio]{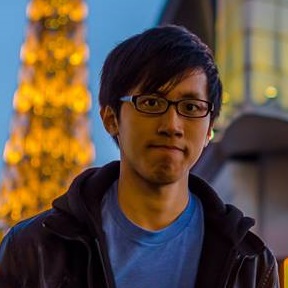}}]{Hao Fei CHENG}
received the bachelor's degree in computer science and engineering from the Department of Computer Science and Engineering at the Hong Kong University of Science and Technology, Hong Kong, in 2015. He is currently working toward the PhD degree in the Department of Computer Science and Engineering at University of Minnesota, U.S.. His main research interests include the areas of social computing and human-computer interaction. 
\end{IEEEbiography}


\begin{IEEEbiography}[{\includegraphics[width=1in,height=1in,clip,keepaspectratio]{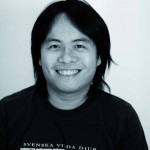}}]{Pan HUI}
has been the Nokia Chair of Data Science and a professor of computer science at the University of Helsinki since September 2017. He also has been a faculty member of the Department of Computer Science and Engineering at the Hong Kong University of Science and Technology since 2013 and an adjunct professor of social computing and networking at Aalto University Finland since 2012. He was a senior research scientist
and then a Distinguished Scientist for Telekom Innovation Laboratories (T-labs) Germany from 2008 to 2015. He also worked for Intel Research Cambridge and Thomson Research Paris while he was pursuing his PhD at the University of Cambridge. He is an IEEE Fellow, an associate editor for IEEE Transactions on Mobile Computing and IEEE Transactions on Cloud Computing, a guest editor for the IEEE Communication Magazine, and an ACM Distinguished
Scientist.
\end{IEEEbiography}




\end{document}